\pdfoutput=1
\documentclass[twocolumn,english,aps,pra,superscriptaddress]{revtex4-1}
\usepackage[T1]{fontenc}
\usepackage[latin9]{inputenc}
\usepackage{amsmath}
\usepackage{amssymb}
\usepackage{graphicx}

\makeatletter
\usepackage[usenames,dvipsnames]{xcolor}
\usepackage[colorlinks=true,citecolor = MidnightBlue,
urlcolor = ForestGreen, linkcolor=BrickRed]{hyperref}
\usepackage{babel}

\makeatother

\usepackage{babel}
\begin{document}

\title{Photon echo using imperfect X-ray pulse with phase fluctuation}

\author{Jinfu Chen}

\address{Beijing Computational Science Research Center, Beijing 100193, China}

\address{Graduate School of China Academy of Engineering Physics, No. 10 Xibeiwang
East Road, Haidian District, Beijing, 100193, China}

\author{Hui Dong}
\email{hdong@gscaep.ac.cn}

\address{Graduate School of China Academy of Engineering Physics, No. 10 Xibeiwang
East Road, Haidian District, Beijing, 100193, China}

\author{Chang-Pu Sun}

\address{Beijing Computational Science Research Center, Beijing 100193, China}

\address{Graduate School of China Academy of Engineering Physics, No. 10 Xibeiwang
East Road, Haidian District, Beijing, 100193, China}
\begin{abstract}
We study the impact of inter-pulse phase fluctuation in free-electron
X-ray laser on the signal in the photon echo spectroscopy, which is
one of the simplest non-linear spectroscopic methods. A two-pulse
echo model is considered with two-level atoms as the sample. The effect
of both fluctuation amplitude and correlation strength of the random
phase fluctuation is studied both numerically and analytically. We
show that the random phase effect only affects the amplitude of the
photon echo, yet not change the recovering time. Such random phase
induces the fluctuation of recovering amplitude in the photon echo
signals among different measurements. We show the normal method of
measuring coherence time retains by averaging across the signals in
different repeats in current paper. 
\end{abstract}
\maketitle

\section{Introduction}

The recent development of x-ray source, especially the large facility
x-ray free-electron laser(XFEL), has attracted vast amount of attentions
\citep{Ullrich2012Annu.Rev.Phys.Chem._63_635--660,Geloni2017,Bostedt2016ReviewsofModernPhysics_88_}
towards detecting properties beyond the scope of traditional instruments.
The unique features of the high brightness, short pulse duration,
and frequency range of XFEL light source open new era in the scientific
investigations in atomic, molecular physics and biology \citep{Lindau1997Nucl.Instrum.MethodsPhys.Res.Sect.A_398_65--68,Prat2015PhysicalReviewSpecialTopics-AcceleratorsandBeams_18_,Bostedt2016ReviewsofModernPhysics_88_}.
One potential application is the implementation of nonlinear spectroscopy
\citep{Mukamel2005Phys.Rev.B_72_,Schweigert2007Phys.Rev.Lett._99_,Schweigert2008Phys.Rev.A_78_,Mukamel2013AnnualReviewofPhysicalChemistry_64_101--127,Healion2013StructuralDynamics_1_014101,Bennett2016PhysicaScripta_T169_014002}
to investigate the dynamics of matter in extreme conditions. The non-linear
spectroscopy typically requires high degree of temporal coherence
\citep{mukamelbook,Schlau-Cohen2011Chem.Phys._386_1--22}, i.e. inter-pulse
phase stability as well as intra-pulse stability \citep{Schlau-Cohen2011Chem.Phys._386_1--22}.
However, pulses generated from many current facilities, may not fulfill
such requirement due to its inter-pulse phase fluctuation \citep{Yu1991PhysicalReviewA_44_5178--5193,Wang1986Nucl.Instrum.MethodsPhys.Res.Sect.A_250_484--489,Bostedt2016ReviewsofModernPhysics_88_,Lee2012Opt.Express_20_9790,yu_high-gain_2000}.
A direct question is how such phase fluctuation affects on the actually
signal, especially on methods of extracting key parameters, e.g coherence
time.

We will investigate the impact of inter-pulse phase fluctuation of
x-ray pulses on photon echo, which is one of the simplest nonlinear
spectroscopy methods, yet fundamental to many advanced spectroscopic
methods, e.g. two-dimensional electronic spectroscopy and two-dimensional
vibrational spectroscopy \citep{Schlau-Cohen2011Chem.Phys._386_1--22,mukamelbook}.
Photon echo \citep{PhysRev.141.391,Shvydko2016Phys.Rev.Lett._116_080801,PhysRev.171.213}
is an optical analogy to spin echo, and is designed to remove ensemble
average for measuring properties of individual spins while maintaining
signal amplitude by avoiding measuring individuals directly \citep{mukamelbook}.
Taking a simple two-level system as an example, an excitation pulse
creates an initial state $\left|\psi\left(0\right)\right\rangle =\alpha\left|g\right\rangle +\beta\left|e\right\rangle $,
where $\left|g\right\rangle $ and $\left|e\right\rangle $ are the
ground and excited state with energies 0 and $\epsilon_{e}$ respectively.
The free evolution brings the system to the state $\left|\psi\left(t\right)\right\rangle =\alpha\left|g\right\rangle +\beta\exp(-i\epsilon_{e}t)\left|e\right\rangle $.
A subsequent $\pi/2$ pulse reverses the population $\left|\psi'\left(t\right)\right\rangle =\alpha\left|e\right\rangle +\beta\exp(-i\epsilon_{e}t)\left|g\right\rangle $.
The later evolution compensates the phase accumulated during the evolution
of ensemble between the two pulses, namely, $\left|\psi\left(T,t\right)\right\rangle =\alpha\exp(-i\epsilon_{e}T)\left|e\right\rangle +\beta\exp(-i\epsilon_{e}t)\left|g\right\rangle .$
At a revival time $T=t$ , the impact of disorder (inhomogenerity)
over the signal is essentially removed. However, it is usually not
easy to achieve the $\pi/2$ pulse due to the weak pulse intensity
in the optical region. One solution is to use non-colinear incident
pulses in order to separate the echo signal from other signal via
phase matching method, which is frequently adopted in non-linear spectroscopy
studies \citep{mukamelbook}.

This paper is organized as follows. In Sec. II, we show the general
model of measuring the signal of two-pulse photon echo on the ensemble
of two-level atoms with imperfect x-ray pulse. In Sec. III, we show
the analytical result of photon echo under the influence of phase
instability along with exact numerical results.

\section{Photon echo with imperfect X-ray pulse}

In the current paper, we consider an ensemble of two-level atoms with
the ground state $\left|g\right\rangle $ and the excited state $\left|e\right\rangle $.
The free Hamiltonian for the two-level atom is 
\begin{equation}
H_{0}=\epsilon_{e}\left|e\right\rangle \left\langle e\right|,
\end{equation}
where we have set the energy of the ground state as $\epsilon_{g}=0$.
The energy levels here are inner-shell electronic states \citep{Mukamel2013AnnualReviewofPhysicalChemistry_64_101--127},
accessible with the frequency of XFEL. The interaction Hamiltonian
between pulses and the atom is given by the dipole interaction $H_{I}=-\vec{\mu}\cdot\vec{E}\left(t\right),$
where $\vec{\mu}$ is the transition dipole and $\vec{E}\left(t\right)$
is the electric field of the incident X-ray pulse. Under the rotation
wave approximation, the interaction Hamiltonian for a pulse with central
frequency $\text{\ensuremath{\nu}}_{0}$ and wave vector $\vec{k}$
is simplified as 
\begin{equation}
H_{I}=-\Omega\left(t\right)e^{-i\nu_{0}t-i\phi\left(t\right)+i\vec{k}\cdot\vec{r}}\left|e\right\rangle \left\langle g\right|+h.c.,
\end{equation}
where $\phi\left(t\right)$ characterizes the random phase of the
X-ray pulse, $\vec{r}$ is the spatial location of the atom, and $\Omega(t)$
is the Rabi frequency. We simplify the model with the square pulse
approximation: the strength of the pulse is a constant $\Omega$ in
the duration for the pulse and diminishes when the pulse ends.

\begin{figure}
\includegraphics{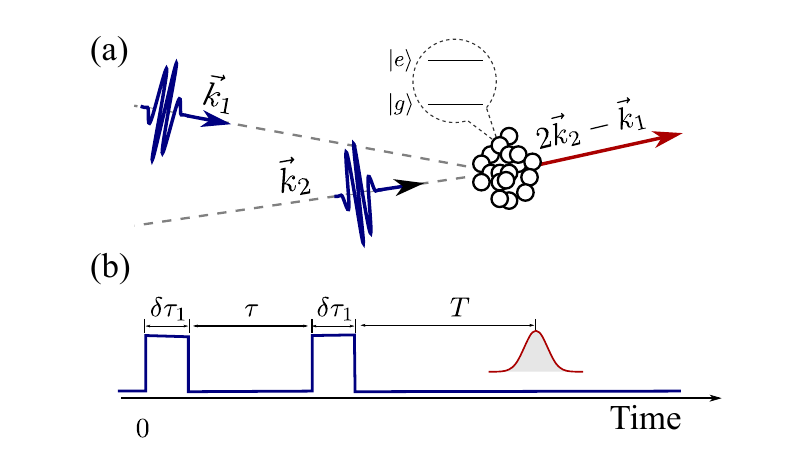}

\caption{The model and pulse sequence. (a) The two incident pulses are directed
to the sample in a non-colinear geometry along two directions $\vec{k}_{1}$
and $\vec{k}_{2}$. The emission of the two-pulse photon echo is along
the direction $2\vec{k}_{2}-\vec{k}_{1}$. The sample is an ensemble
of two-level atoms with the ground state $\left|g\right\rangle $
and excited state $\left|e\right\rangle $. (b) The blue line shows
the two pulses, and the red line show the signal. The duration of
the two pulse are $\delta\tau_{1}$ and $\delta\tau_{2}$, and the
delay time is $\tau$. And the measurement of the echo signal acts
at $T$ time after the end of the second pulse.}
\end{figure}

Here, we consider the two-pulse photon echo, and the two pulses are
set to be resonated to the atom $\nu_{0}=\epsilon_{e}.$ The first
pulse interacts with the atoms with the duration $\delta\tau_{1}$,
while the second pulse interacts with the atoms with the duration
time $\delta\tau_{2}$ after delay time $\tau$ of the end of the
first pulse. The evolution matrices for each pulse are $U_{i}(\vec{k}_{i},\delta\tau_{i})\,(i=1,2)$,
and the free evolution of the atom is $U_{0}(t_{f},t_{i})=U_{0}(t_{f}-t_{i})=\exp[-iH_{0}(t_{f}-t_{i})]$,
where $t_{i}(t_{f})$ is the initial (final) time of the free evolution.
The final wave function of the atom at delay time $T$ is obtained
as

\begin{equation}
\left|\psi\left(T,\tau\right)\right\rangle =U_{0}\left(T\right)U_{2}(\vec{k}_{2},\delta\tau_{2})U_{0}\left(\tau\right)U_{1}(\vec{k}_{1},\delta\tau_{1})\left|\psi_{0}\right\rangle ,\label{eq:5}
\end{equation}
where the initial state $\left|\psi_{0}\right\rangle $ is usually
considered as the ground state $\left|\psi_{0}\right\rangle =\left|g\right\rangle $.
To derive the evolution matrices $U_{i}(k_{i},\delta\tau_{i})$ for
each pulse, we rewrite the Hamiltonian in the interacting picture
as
\begin{equation}
\hat{H}_{I}(t)=-\left(\begin{array}{cc}
0 & \hbar\Omega_{a}e^{i\phi(t)-i\vec{k}\cdot\vec{r}}\\
\hbar\Omega_{a}e^{-i\phi(t)+i\vec{k}\cdot\vec{r}} & 0
\end{array}\right).\label{eq:hamiltonian}
\end{equation}
The time dependence of Eq. (\ref{eq:hamiltonian}) only comes from
the random phase factor $\phi\left(t\right)$. With the following
definitions
\begin{equation}
H_{1}\left(\vec{k}\right)=\left(\begin{array}{cc}
0 & \hbar\Omega e^{-i\vec{k}\cdot\vec{r}}\\
\hbar\Omega e^{i\vec{k}\cdot\vec{r}} & 0
\end{array}\right),
\end{equation}

\begin{equation}
H_{2}\left(\vec{k}\right)=\left(\begin{array}{cc}
0 & -i\hbar\Omega e^{-i\vec{k}\cdot\vec{r}}\\
i\hbar\Omega e^{i\vec{k}\cdot\vec{r}} & 0
\end{array}\right),
\end{equation}
\begin{equation}
H_{3}\left(\vec{k}\right)=\left(\begin{array}{cc}
\hbar\Omega & 0\\
0 & -\hbar\Omega
\end{array}\right),
\end{equation}
$\hat{H}_{I}(t)$ is rewritten in a compact form

\begin{equation}
\hat{H}_{I}(t)=-\cos\left[\phi(t)\right]H_{1}+\sin\left[\phi(t)\right]H_{2}.\label{eq:5-1}
\end{equation}
The three operators $H_{l}\,(l=1,2,3)$ satisfy the commutation relation
of angular momentum operators $[H_{i}(\vec{k}),H_{j}(\vec{k})]=2i\hbar\sum_{l=1}^{3}\Omega H_{l}(\vec{k})\epsilon_{ijl}.$
Following the Wei-Norman algebra method \citep{wei_lie_1963,sun_wei-norman_1991},
the evolution matrix for a pulse is written as
\begin{equation}
\hat{U}_{\vec{k},\phi}\left(t,0\right)=e^{-i\chi_{3}(t)H_{3}}e^{-i\chi_{2}(t)H_{2}}e^{-i\chi_{1}(t)H_{1}},\label{eq:9}
\end{equation}
where $\vec{k}$ is the wave vector and $\phi$ is a certain realization
of the random phase. With the commutation relation of $H_{l}\,(l=1,2,3)$,
we have derived the differential equations for the time-dependent
parameters $\{\chi_{l}\left(t\right)\},\,(l=1,2,3)$ in Appendix A

\begin{equation}
\left\{ \begin{array}{l}
\dot{\chi}_{3}=-\cos(\phi+2\chi_{3}\Omega)\tan2\chi_{2}\Omega\\
\dot{\chi}_{2}=\sin(\phi+2\chi_{3}\Omega)\\
\dot{\chi}_{1}=-\cos(\phi+2\chi_{3}\Omega)\sec2\chi_{2}\Omega.
\end{array}\right.\label{eq:116}
\end{equation}
The initial condition is $\chi_{l}\left(0\right)=0,\;l=1,2,3$. The
non-linear time-dependent differential equations (\ref{eq:116}) are
accessible to be solved numerically with a given $\phi(t)$. 

Next, we change the evolution matrices derived by Eq.(\ref{eq:9})
from the interacting picture to Schrodinger picture, which is linked
by a free evolution $U_{1}(\vec{k}_{1},\delta\tau_{1})=U_{0}\left(\delta\tau_{1}\right)\hat{U}_{\vec{k}_{1},\phi_{1}}\left(\delta\tau_{1},0\right)$
and $U_{2}(\vec{k}_{2},\delta\tau_{2})=U_{0}\left(\delta\tau_{2}\right)\hat{U}_{\vec{k_{2}},\phi_{2}}\left(\delta\tau_{2},0\right).$
$U_{0}(\delta\tau_{i})$ is absorbed to the free evolution part or
neglected when $\delta\tau_{i}$ is small compared to the interval
time $\tau$ and the measurement time $T$. Combined with Eq. (\ref{eq:5}),
the echo term is derived by sorting terms with the phase factor matching
$\exp[i(2\vec{k}_{2}-\vec{k}_{1})\cdot\vec{r}]$ as follows,

\begin{equation}
\begin{array}{l}
\left\langle \psi\left(T,\tau\right)\right|\mu_{t}\left|\psi\left(T,\tau\right)\right\rangle \\
\sim i\mu^{*}e^{-i\left(T-\tau\right)\epsilon_{e}}e^{i(2\vec{k}_{2}-\vec{k}_{1})\cdot\vec{r}}\\
\times\frac{e^{2i\Omega\left(\zeta_{3}-\chi_{3}\right)}}{8}\left[\sin((\zeta_{2}+\zeta_{1})\Omega)+i\sin((\zeta_{2}-\zeta_{1})\Omega)\right]{}^{2}\\
\times\left[\sin(2(\chi_{2}-\chi_{1})\Omega)+2i\sin(2\chi_{1}\Omega)+\sin(2(\chi_{2}+\chi_{1})\Omega)\right].
\end{array}\label{eq:17}
\end{equation}

For the ensemble of atoms, their energy $\epsilon_{e}$ between the
ground state and the excited state has fluctuations, assumed as Gaussian
distribution with mean value $\epsilon_{0}$ and variance $\sigma_{0}^{2}$
with the following form,
\begin{equation}
p\left(\epsilon_{e}\right)=\frac{1}{\sqrt{2\pi}\sigma_{0}}\exp\left[-\frac{(\epsilon_{e}-\epsilon_{0})^{2}}{2\sigma_{0}^{2}}\right].
\end{equation}
The summation over transition energies of different molecules contributes
a Gaussian decay with $(T-\tau)$, namely,

\begin{equation}
\sum_{\epsilon_{e}}e^{-i\left(T-\tau\right)\epsilon_{e}}\rightarrow e^{-\frac{1}{2}\sigma_{0}^{2}(T-\tau)^{2}-i\left(T-\tau\right)\epsilon_{0}}.
\end{equation}
At the revival time $T=\tau$, the average over different molecules
vanishes so that decoherence time can be directly detected. The amplitude
$\mathcal{A}$ of the photon echo signal is the square of the absolute
value of Eq. (\ref{eq:17})

\begin{equation}
\begin{array}{l}
\mathcal{A}=\frac{\left|\mu\right|^{2}e^{-\sigma_{0}^{2}(T-\tau)^{2}}}{64}\left[\sin^{2}((\zeta_{2}+\zeta_{1})\Omega)+\sin^{2}(\zeta_{2}-\zeta_{1})\Omega\right]^{2}\\
\left[\left(\sin(2(\chi_{2}-\chi_{1})\Omega)+\sin(2(\chi_{2}+\chi_{1})\Omega)\right)^{2}+4\sin^{2}(2\chi_{1}\Omega)\right]
\end{array}
\end{equation}

For the ideal case with no random phase ($\phi\left(t\right)=\mathrm{constant}$),
we obtain the amplitude
\begin{equation}
\mathcal{A}_{\mathrm{ideal}}=\frac{\left|\mu\right|^{2}e^{-\sigma_{0}^{2}(T-\tau)^{2}}}{4}\sin^{4}(\Omega\delta\tau_{2})\sin^{2}(2\Omega\delta\tau_{1}).
\end{equation}
 with $\chi_{2}^{(0)}=\chi_{3}^{(0)}=\zeta_{2}^{(0)}=\zeta_{3}^{(0)}=0,\chi_{1}^{(0)}=-\delta\tau_{1},\zeta_{1}^{(0)}=-\delta\tau_{2}$.
It is clear that the random phase only affects the amplitude of the
photon echo. A factor $\text{\ensuremath{\mathcal{F}} }$is defined
to represent the value of the amplitude 
\begin{align}
\mathcal{F} & =\{\sin^{2}[(\zeta_{2}+\zeta_{1})\Omega]+\sin^{2}[(\zeta_{2}-\zeta_{1})\Omega]\}^{2}\nonumber \\
 & \times\left[[\sin[2(\chi_{2}-\chi_{1})\Omega]+\sin[2(\chi_{2}+\chi_{1})\Omega]]^{2}+4\sin(2\chi_{1}\Omega)^{2}\right].\label{eq:19-1}
\end{align}
In the following discussion, we consider the two pulses is the same
except different direction, namely, $\delta\tau_{1}=\delta\tau_{2}=\delta\tau$,
and $\chi_{i}=\zeta_{i}$ $\left(i=1,2,3\right).$ For the case without
phase fluctuation ($\phi\left(t\right)=\mathrm{constant}$), the factor
$\mathcal{F}$ is simply 
\begin{equation}
\mathcal{F}_{\mathrm{ideal}}=16\sin^{4}(\Omega\delta\tau)\sin^{2}\left(2\Omega\delta\tau\right).
\end{equation}

\begin{figure}
\includegraphics[scale=0.42]{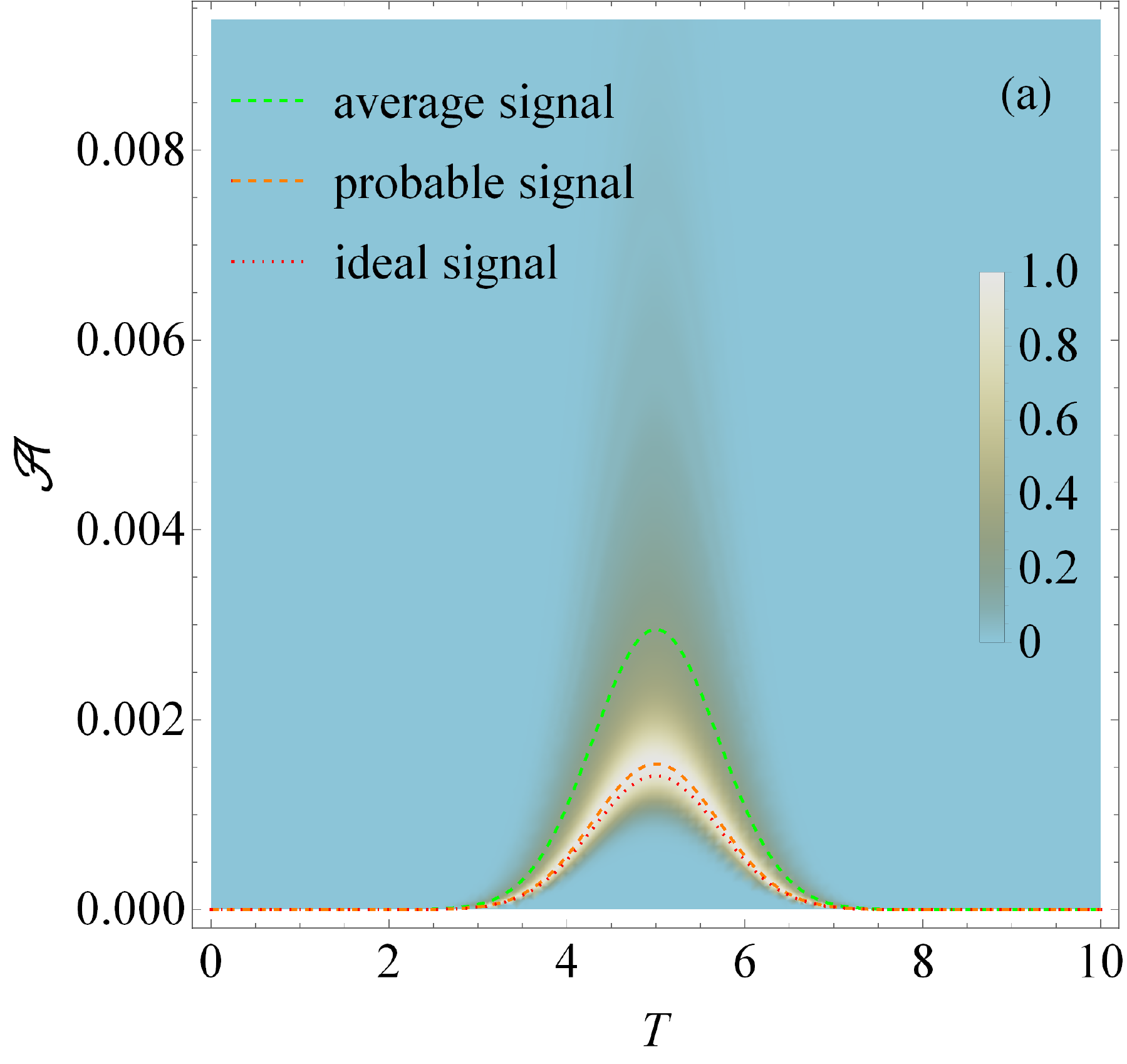}

\includegraphics[scale=0.42]{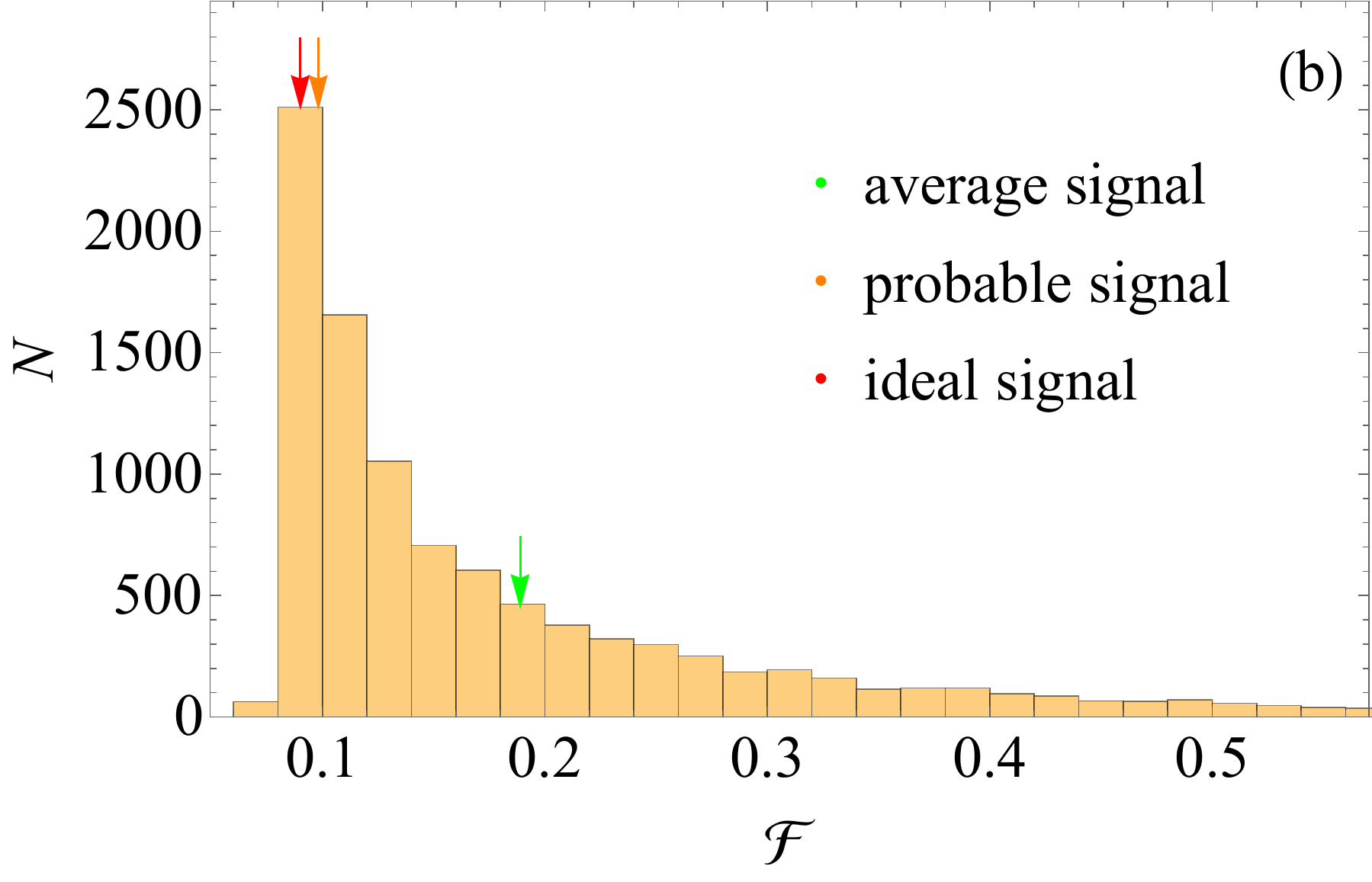}\caption{The photon echo signal and the distribution of strength factor $\mathcal{F}$.
The parameters are chosen as $\tau=5$, $\sigma_{0}=1$,$\Omega=1$,
$\delta\tau=4.75$ ,$\gamma=1/4.587$, $\Phi=0.08$. (a) the distribution
of the signal at time $T$. The color shows the probability at given
time $T$ with the amplitude $\mathcal{A}$. The green curve and the
orange curve shows the average signal and the most probable signal
respectively, while the red curve shows the ideal echo signal without
any phase randomness. (b) the distribution of the strength factor
$\mathcal{F}$ with 10000 repeats for echo signal at $T=\tau$. The
arrows show the average, most probable, and ideal signal with the
same color scheme as in subfigure (a).}
\end{figure}

In Fig 2(a), we show the distribution of the signal intensity $\mathcal{A}$
as a function of $T$. The random phase elicits fluctuation to the
signal and enlarges the average value. In the simulation, we generate
the random function $\phi(t)$ with the Ornstein-Uhlenbeck process.
The average of $\phi(t)$ is zero $\left\langle \phi\left(t\right)\right\rangle =0$,
and its two-point correlation function satisfies
\begin{equation}
\left\langle \phi\left(t_{1}\right)\phi\left(t_{2}\right)\right\rangle =\Phi^{2}e^{-\gamma\left|t_{1}-t_{2}\right|},
\end{equation}
where $\Phi$ is the fluctuation amplitude of the random phase and
$\gamma$ is the correlation strength. The amplitude of signal is
evaluated via Eq. (\ref{eq:17}) with $\chi_{i}$ and $\zeta_{i}$,
which is numerically solved the differential equation (\ref{eq:116}).
The statistics is calculated with 10000 repeats of the current process
by generating different random function $\phi(t)$. In the simulation,
we have chosen parameters as follows, $\tau=5,\:\sigma=1,\:\Omega=1,\:\delta\tau=4.75,\:\gamma=1/4.587,$
and $\Phi=0.08$. In Fig. 2(a), we show the average signal with green
dashed line, the most probable signal with orange dashed line, and
the signal without random phase with red dotted line. We further show
the randomness of the factor $\mathcal{F}$ in Fig 2(b), whose distribution
$p(\mathcal{F})$ is not Gaussian. The mean value of $\mathcal{F}$
is larger than the most probable case and the ideal case (the case
without random phase). It is clear that the random phase induces fluctuation
on the strength of the signal of the photon echo.

With the observation of the randomness of the echo amplitude, it is
meaningful to calculate the average signal with different repeats.
Here, we try to derive perturbation results for the average amplitude
$\langle\mathcal{A}\rangle$ with random phase. We consider the random
phase is small and apply the approximation $\cos\phi(t)\approx1,\:\sin\phi(t)\approx\phi(t)$
to obtain the linear differential equation of Eq.(\ref{eq:116}) for
$\chi_{2}$ and $\chi_{3}$ as follows. The differential equation
for $\chi_{1}$ is kept for second order to obtain the signal amplitude
to the second order

\begin{equation}
\left\{ \begin{array}{l}
\dot{\chi}_{3}=-2\Omega\chi_{2}\\
\dot{\chi}_{2}=\phi+2\Omega\chi_{3}\\
\dot{\chi}_{1}=-1+\frac{1}{2}\left(\phi+2\chi_{3}\Omega\right)^{2}-\frac{1}{2}\left(2\chi_{2}\Omega\right)^{2}.
\end{array}\right.\label{eq:117}
\end{equation}
Now, the current equation (\ref{eq:117}) has an integral solution
\begin{equation}
\begin{array}{c}
\chi_{3}\left(t\right)=-\int_{0}^{t}\sin2\Omega\left(t-t_{1}\right)\phi\left(t_{1}\right)dt_{1}\\
\chi_{2}\left(t\right)=\int_{0}^{t}\cos2\Omega\left(t-t_{1}\right)\phi\left(t_{1}\right)dt_{1}\\
\chi_{1}\left(t\right)=-t+\int_{0}^{t}\frac{1}{2}\phi^{2}+2\Omega\chi_{3}\phi+2\Omega^{2}\chi_{3}^{2}-2\Omega^{2}\chi_{2}^{2}dt_{1}.
\end{array}\label{eq:121}
\end{equation}
For small random phase $\phi(t)$, we expand the factors $\chi_{i},\:i=1,2,3$
to their first order $\chi_{i}=\chi_{i}^{(0)}+\chi_{i}^{(1)}$ , where
$\chi_{i}^{(0)}$ is the average value, i.e., $\chi_{1}^{(0)}=0,\:\chi_{2}^{(0)}=0,\,\chi_{3}^{(0)}=-\delta\tau$
and $\chi_{i}^{(1)}$ gives the fluctuation due to the random phase.
We obtain the explicit form of the factor $\mathcal{F}$ under the
perturbation formalism

\begin{equation}
\begin{array}{l}
\begin{array}{l}
\mathcal{F}=\mathcal{F}_{\mathrm{ideal}}-128\chi_{1}^{(1)}\Omega\sin^{5}(\delta\tau\Omega)(2\cos(\delta\tau\Omega)+\cos(3\delta\tau\Omega))\end{array}\\
+64\left(\chi_{2}^{(1)}\right)^{2}\Omega^{2}\sin^{4}(\delta\tau\Omega)\cos(2\delta\tau\Omega)(2\cos(2\delta\tau\Omega)+1).
\end{array}\label{eq:21}
\end{equation}
Eq. (\ref{eq:21}) contains $\chi_{2}^{(1)}$ to the second order
and $\chi_{1}^{(1)}$ to the first order. It is verified numerically
that $\chi_{1}^{(1)}$ is as the same order as $(\chi_{2}^{(1)})^{2}$
in Appendix B. And they are both the lowest order contributing to
$\mathcal{F}$.

The factor $\mathcal{F}$ is a random variable due to the random phase
$\phi(t)$. We derive the average value $\left\langle (\chi_{2}^{(1)})^{2}\right\rangle $
and $\left\langle \chi_{1}^{(1)}\right\rangle $ by the two-point
correlation function. The detailed calculation is attached in the
Appendix B. The analytical result for the mean value of $\mathcal{F}$
becomes

\begin{equation}
\begin{array}{l}
\mathcal{\left\langle \mathcal{F}\right\rangle }=\mathcal{F}_{\mathrm{ideal}}\\
+64\Omega^{2}\sin^{4}(\delta\tau\Omega)\cos(2\delta\tau\Omega)(2\cos(2\delta\tau\Omega)+1)\left\langle (\chi_{2}^{(1)})^{2}\right\rangle \\
-128\Omega\sin^{5}(\delta\tau\Omega)(2\cos(\delta\tau\Omega)+\cos(3\delta\tau\Omega))\left\langle \chi_{1}^{(1)}\right\rangle .
\end{array}\label{eq:29}
\end{equation}

\begin{figure}[h]
\includegraphics[width=7.5cm]{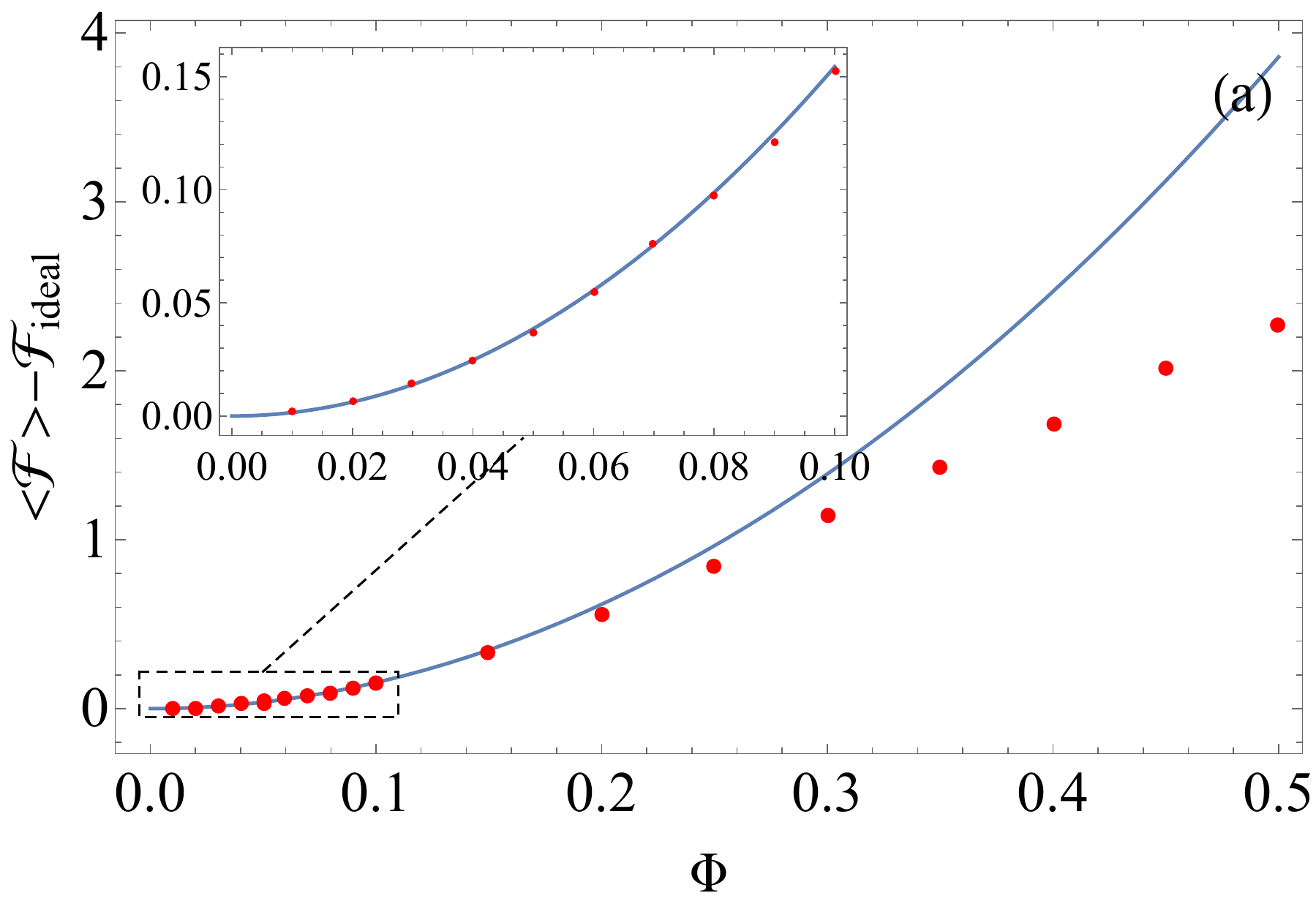}

\includegraphics[width=7.5cm]{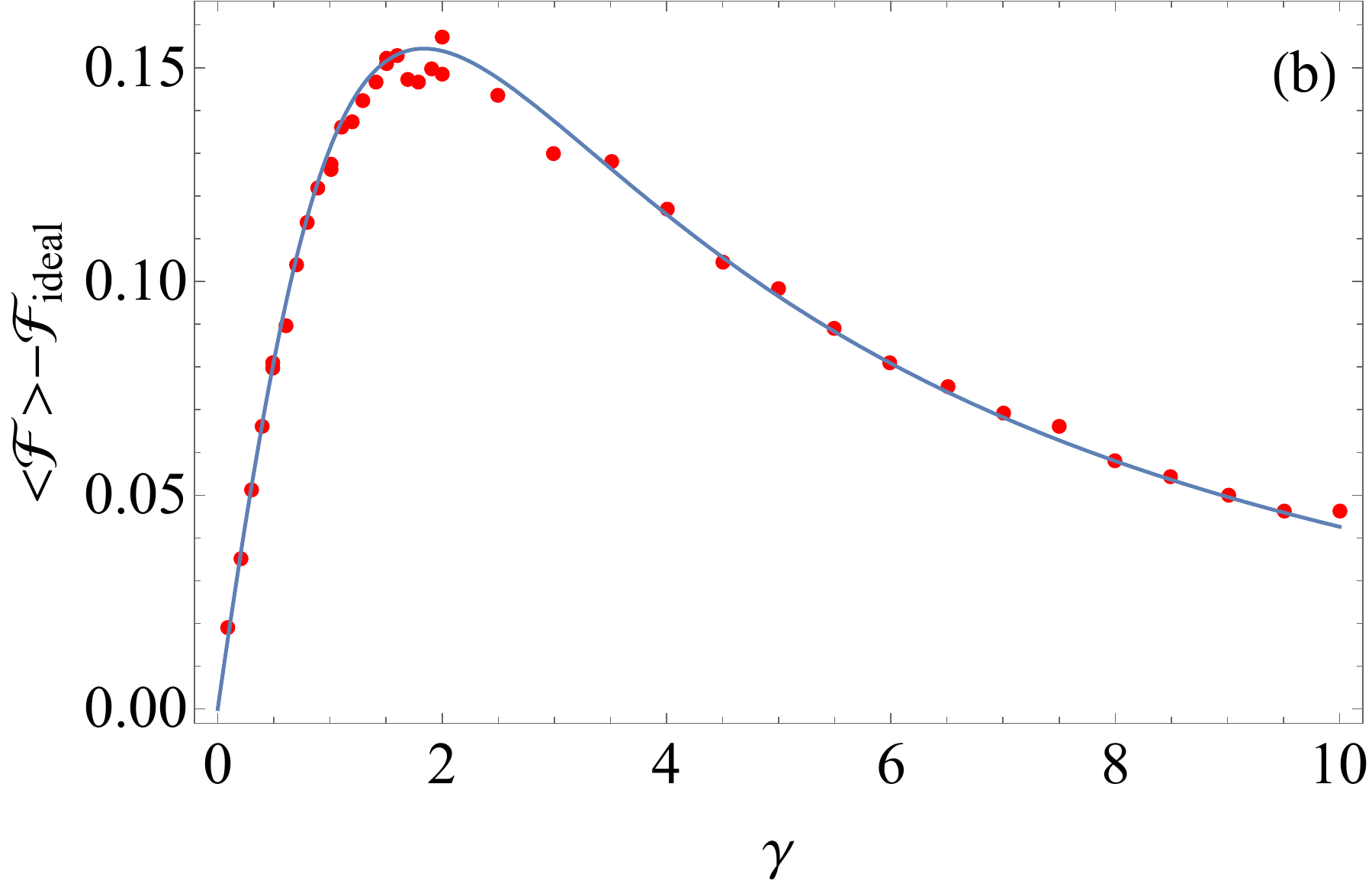}

\caption{The relation between $\left\langle \mathcal{F}\right\rangle -\mathcal{F}_{\mathrm{ideal}}$
and the fluctuation amplitude $\Phi$ and the correlation strength
$\gamma$ . The parameters are chosen as $\Omega=1$, $\delta\tau=4.75$
,$\gamma=1/4.587$ for the figure 3 (a), (c), $\Phi=0.05$ for figure
3 (b). The red points shows the numerical calculation and the solid
curve shows the analytical result.}
\end{figure}

In Fig 3 (a), we plot the average signal $\left\langle \mathcal{F}\right\rangle -\mathcal{F}_{\mathrm{ideal}}$
as a function of the fluctuation amplitude $\Phi$ with the correlation
strength $\gamma$ fixed. In the figure, red dots show the exact result
by numerical calculation, and lines represent the analytical result
in Eq. (\ref{eq:29}). For small fluctuation amplitude $\Phi$, the
analytical result matches numerical calculation well, as illustrated
in subset of Fig 3 (a). However, the analytical result deviates from
exact numerical result for large $\Phi$, e.g. $\Phi>0.3$. In Fig
3 (b), we plot the average signal $\left\langle \mathcal{F}\right\rangle -\mathcal{F}_{\mathrm{ideal}}$
as a function of the correlation strength $\gamma$ with the fluctuation
amplitude $\Phi$ fixed. The analytical result matches numerical calculation
well whether for large or small $\gamma$.

In above discussion, we have shown the random phase effect on the
signal amplitude of an ensemble of two-level atoms without any decoherence.
The key function of photon echo is to measure the decoherence time
$\tau_{c}$. In open systems, the environment induces a decoherence
to the atoms, which contributes to the decreasing of the non-diagonal
term $\rho_{eg}(t)=\exp[-i\epsilon_{e}t-t/\tau_{c}]$. With the decoherence
effect, the signal derived in Eq. (16) becomes

\begin{equation}
\begin{array}{l}
\mathcal{A}_{\mathrm{open}}=\frac{\left|\mu\right|^{2}e^{-\sigma_{0}^{2}(T-\tau)^{2}-\left(T+\tau\right)/\tau_{c}}}{64}\mathcal{F}.\end{array}
\end{equation}
At the revival time $T=\tau$, the average signal is 
\begin{equation}
\langle\mathcal{A}_{\mathrm{open}}\rangle=\left|\mu\right|^{2}\exp(-2\tau/\tau_{c})\langle\mathcal{F}\rangle.\label{eq:26}
\end{equation}
With fixed $\delta\tau$ and given random phase, the average for the
factor $\mathcal{F}$ is invariant. To measure the coherence time
$\tau_{c}$, we still follow the normal way of changing the delay
time $\tau$ and obtain the signal amplitude at $T=\tau$. By taking
average over different repeats, the coherence time is recovered via
Eq. (\ref{eq:26}).

Currently, the experimental setup of x-ray photon echo is achievable
with the split-delay approach \citep{Gutt2008Opt.Express_17_55,Roseker2011J.SynchrotronRad._18_481--491},
where the x-ray pulse is split by a silicon beam splitter\citep{Roseker2011J.SynchrotronRad._18_481--491}.
The change to the setup in Ref \citep{Gutt2008Opt.Express_17_55}
is to direct the splitted two pulses to the sample along two directions.
With the split-delay approach, the phase difference between pulses
is fixed with delay time. And the phase fluctuation of each pulses
is theoretically considered in the current paper.

\section{Conclussion}

We have theoretically calculated the impact of phase randomness on
the photon echo experiment, which is fundamental to many other non-linear
spectroscopy, such as two-dimensional spectroscopy. We found that
the phase randomness will induce fluctuation in the photon echo signal,
yet not affect the rephasing time. By averaging the signal from different
repeats, the normal way of photon echo is still effective for measuring
the decoherence time.
\begin{acknowledgments}
This work is supported by NSFC (Grants No. 11421063, No. 11534002),
the National Basic Research Program of China (Grant No. 2016YFA0301201
\& No. 2014CB921403), and the NSAF (Grant No. U1730449 \& No. U1530401).
\end{acknowledgments}

\bibliographystyle{apsrev4-1}
\bibliography{xraypaper1}

\begin{widetext}

\appendix

\section{Wei Norman Method}
\begin{widetext}
In this appendix, we show the detailed derivation of the differential
equation (\ref{eq:116}). The derivation is based on the Wei-Norman
algebra method\citep{sun_wei-norman_1991,wei_lie_1963}. The differential
of Eq. (\ref{eq:9}) is calculated 
\begin{equation}
\begin{array}{l}
\frac{d}{dt}U_{p,\phi}(t,0)=-i\dot{\chi}_{3}H_{3}U_{p,\phi}(t,0)\\
\qquad-i\dot{\chi}_{2}e^{-i\chi_{3}H_{3}}H_{2}e^{-i\chi_{2}H_{2}}e^{-i\chi_{1}H_{1}}\\
\qquad-i\dot{\chi}_{1}U_{p,\phi}(t,0)H_{1}
\end{array}\label{eq:33}
\end{equation}
With the commutations

\begin{equation}
e^{-i\chi_{3}H_{3}}H_{2}e^{i\chi_{3}H_{3}}=H_{2}\cos2\Omega\chi_{3}-H_{1}\sin2\Omega\chi_{3},
\end{equation}
\begin{equation}
\begin{array}{l}
e^{-i\chi_{3}H_{3}}e^{-i\chi_{2}H_{2}}H_{1}e^{i\chi_{2}H_{2}}e^{i\chi_{3}H_{3}}=-H_{3}\sin2\Omega\chi_{2}\\
\qquad+H_{2}\sin2\Omega\chi_{3}\cos2\Omega\chi_{2}\\
\qquad+H_{1}\cos2\hbar\Omega\chi_{3}\cos2\Omega\chi_{2}
\end{array},
\end{equation}
Eq. (\ref{eq:33}) is rewritten as

\begin{equation}
\begin{array}{l}
i\frac{\partial}{\partial t}U_{p,\phi}\left(t,0\right)=[\left(\dot{\chi}_{3}-\dot{\chi}_{1}\sin2\Omega\chi_{2}\right)H_{3}\\
+\left(\dot{\chi}_{1}\sin2\Omega\chi_{3}\cos2\Omega\chi_{2}+\dot{\chi}_{2}\cos2\Omega\chi_{3}\right)H_{2}\\
+\left(\dot{\chi}_{1}\cos2\Omega\chi_{3}\cos2\Omega\chi_{2}-\dot{\chi}_{2}\sin2\Omega\chi_{3}\right)H_{1}]U_{p,\phi}\left(t,0\right).
\end{array}
\end{equation}
The coefficients must match the Schrodinger equation (\ref{eq:5-1})

\begin{equation}
\left(\begin{array}{ccc}
0 & -\sin2\Omega\chi_{3} & \cos2\Omega\chi_{3}\cos2\Omega\chi_{2}\\
0 & \cos2\Omega\chi_{3} & \sin2\Omega\chi_{3}\cos2\Omega\chi_{2}\\
1 & 0 & -\sin2\Omega\chi_{2}
\end{array}\right)\left(\begin{array}{c}
\dot{\chi}_{3}\\
\dot{\chi}_{2}\\
\dot{\chi}_{1}
\end{array}\right)=\left(\begin{array}{c}
-\cos\phi\\
\sin\phi\\
0
\end{array}\right).
\end{equation}
The differential equations are obtained by taking the inverse matrix

\begin{equation}
\left\{ \begin{array}{l}
\dot{\chi}_{3}=-\cos\phi\cos2\chi_{3}\Omega\tan2\chi_{2}\Omega+\sin\phi\sin2\chi_{3}\Omega\tan2\chi_{2}\Omega\\
\dot{\chi}_{2}=\cos2\chi_{3}\Omega\sin\phi+\cos\phi\sin2\chi_{3}\Omega\\
\dot{\chi}_{1}=-\cos\phi\cos2\chi_{3}\Omega\sec2\chi_{2}\Omega+\sin\phi\sec2\chi_{2}\Omega\sin2\chi_{3}\Omega
\end{array}\right.\label{eq:116-1}
\end{equation}
With further simplification we obtain Eq. (\ref{eq:116}).

\section{The Calculation of $\left\langle \left(\chi_{2}^{(1)}\right)^{2}\right\rangle $
and $\left\langle \chi_{1}^{(1)}\right\rangle $}

Here, we give the detailed calculation for$\left\langle (\chi_{2}^{(1)})^{2}\right\rangle $
and $\left\langle \chi_{1}^{(1)}\right\rangle $. With Eq. (\ref{eq:121}),
we can calculate the average value of $(\chi_{2}^{(1)})^{2}$

\begin{equation}
\left\langle (\chi_{2}^{(1)})^{2}\right\rangle =\int_{0}^{\delta\tau}\int_{0}^{\delta\tau}\cos2\Omega\left(t-t_{1}\right)\cos2\Omega\left(t-t_{2}\right)\left\langle \phi\left(t_{1}\right)\phi\left(t_{2}\right)\right\rangle dt_{2}dt_{1}.\label{eq. chi 22}
\end{equation}
The result of the integral gives 
\begin{equation}
\left\langle (\chi_{2}^{(1)})^{2}\right\rangle =\Phi^{2}\left(\frac{\gamma\delta\tau}{\gamma^{2}+4\Omega^{2}}+\frac{2\gamma e^{-\gamma\delta\tau}\left(\gamma\cos\left(2\delta\tau\Omega\right)-2\Omega\sin\left(2\delta\tau\Omega\right)\right)}{\left(\gamma^{2}+4\Omega^{2}\right)^{2}}+\frac{\gamma\sin\left(4\delta\tau\Omega\right)-2\Omega\cos\left(4\delta\tau\Omega\right)}{4\Omega\left(\gamma^{2}+4\Omega^{2}\right)}+\frac{8\Omega^{2}-6\gamma^{2}}{4\left(\gamma^{2}+4\Omega^{2}\right)^{2}}\right).
\end{equation}

It is similar to calculate the average value of $\chi_{1}^{(1)}$

\begin{equation}
\left\langle \chi_{1}^{(1)}\right\rangle =\left\langle \int_{0}^{\delta\tau}\left(\frac{1}{2}\phi^{2}+2\Omega\chi_{3}\phi+2\Omega^{2}\chi_{3}^{2}-2\Omega^{2}\chi_{2}^{2}\right)dt_{1}\right\rangle .
\end{equation}

The result is 
\begin{equation}
\left\langle \chi_{1}^{(1)}\right\rangle =\Phi^{2}\left(\frac{\gamma^{2}\delta\tau}{2\gamma^{2}+8\Omega^{2}}-\frac{2\gamma\Omega e^{-\gamma\delta\tau}(\gamma\sin(2\delta\tau\Omega)+2\Omega\cos(2\delta\tau\Omega))}{\left(\gamma^{2}+4\Omega^{2}\right)^{2}}+\frac{\Omega\sin(4\delta\tau\Omega)-\gamma\sin^{2}(2\delta\tau\Omega)}{2\left(\gamma^{2}+4\Omega^{2}\right)}+\frac{4\gamma\Omega^{2}}{\left(\gamma^{2}+4\Omega^{2}\right)^{2}}\right).\label{eq:chi1}
\end{equation}
And we show the numerical calculation matches the analytical result
in Figure 5, which shows that $\left\langle (\chi_{2}^{(1)})^{2}\right\rangle $
and $\left\langle \chi_{1}^{(1)}\right\rangle $ are the same order
and should be kept for the perturbation.

\begin{figure*}
\includegraphics[width=7.5cm]{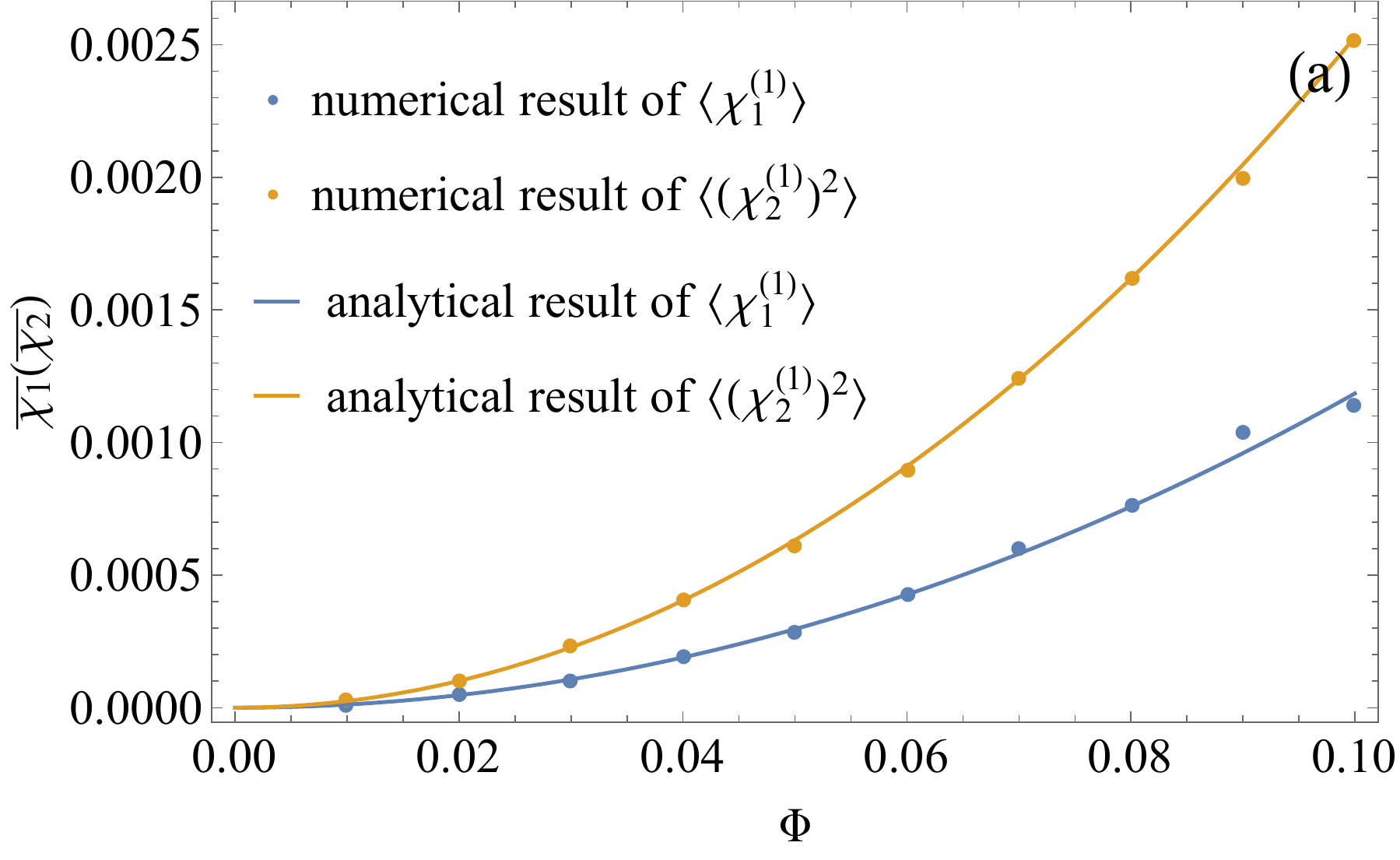}\includegraphics[width=7.5cm]{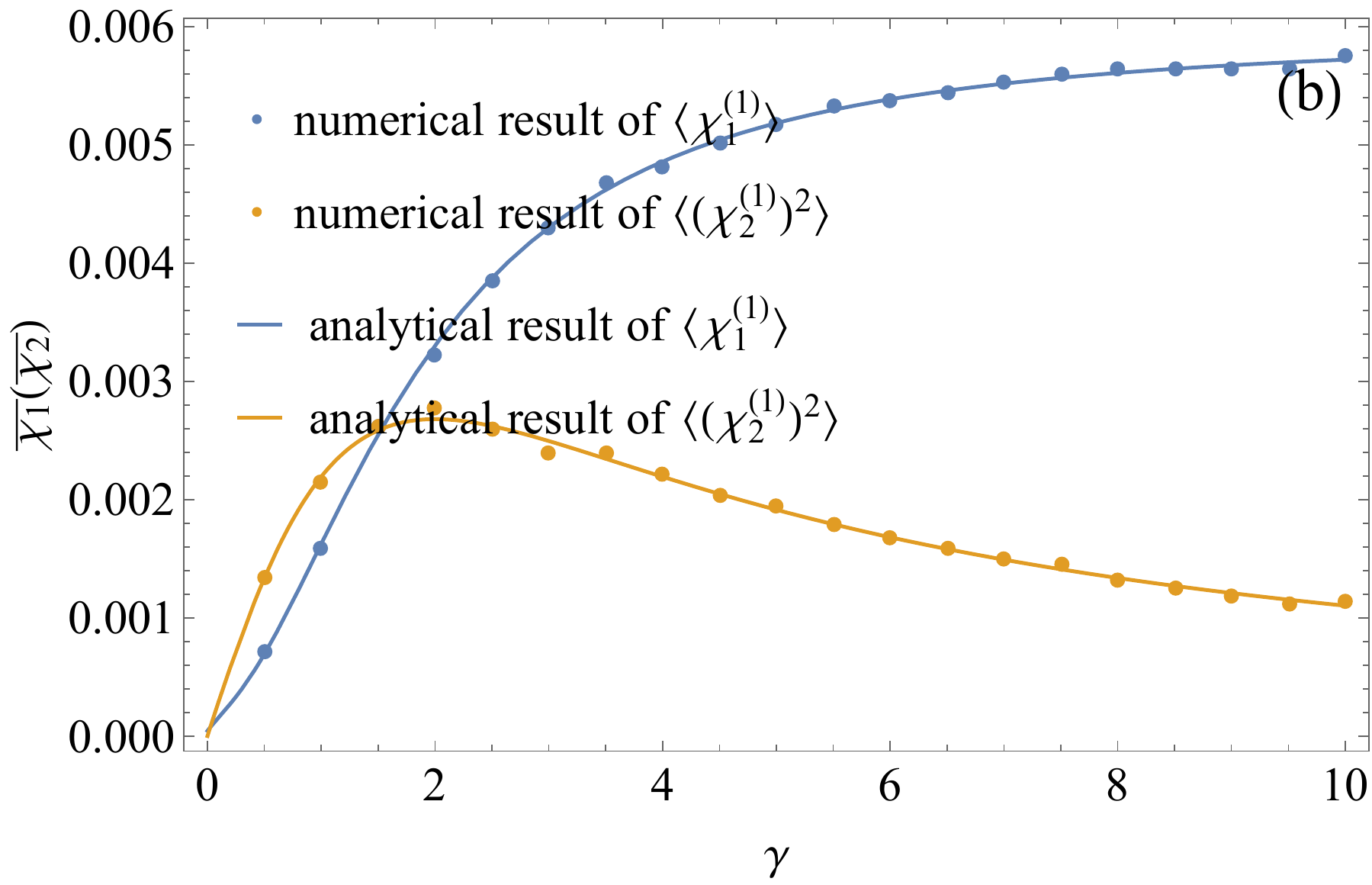}

\caption{The numerical and analytical result of $\left\langle \chi_{1}^{(1)}\right\rangle $
and $\left\langle (\chi_{2}^{(1)})^{2}\right\rangle $ , the point
shows the numerical calculation of the average value, and the line
shows the analytical result.}
\end{figure*}
\end{widetext}

\end{widetext}
\end{document}